\documentclass[12pt]{iopart}

\usepackage{iopams} 
\usepackage{siunitx}
\usepackage{gensymb}
\usepackage{graphicx}
\usepackage{makecell}
\usepackage{subcaption}
\usepackage[section]{placeins}

\DeclareSIUnit\sq{\ensuremath{\Box}}

\begin{document}

\title[Reduced ITO for Transparent Superconducting Electronics]{Reduced ITO for Transparent Superconducting Electronics}

\author{Emma Batson, Marco Colangelo, John Simonaitis, Eyosias Gebremeskel, Owen Medeiros, Mayuran Saravanapavanantham, Vladimir Bulovic, P. Donald Keathley, Karl K. Berggren}

\address{Department of Electrical Engineering and Computer Science, Massachusetts Institute of Technology, Cambridge, MA}
\ead{emmabat@mit.edu}
\vspace{10pt}
\begin{indented}
\item[] December 2022
\end{indented}

\begin{abstract}
Absorption of light in superconducting electronics is a major limitation on the quality of circuit architectures that integrate optical components with superconducting components. A 10 nm thick film of a typical superconducting material like niobium can absorb over half of any incident optical radiation. We propose instead using superconductors which are transparent to the wavelengths used elsewhere in the system. In this paper we investigated reduced indium tin oxide (ITO) as a potential transparent superconductor for electronics.  We fabricated and characterized superconducting wires of reduced indium tin oxide. We also showed that a $\SI{10}{nm}$ thick film of the material would only absorb about 1 - 20\% of light between 500 - 1700 nm.
\end{abstract}

\vspace{2pc}

\submitto{\SUST}
%\maketitle

\section{Introduction}

Integration of photonics on-chip with superconducting electronics is necessary for scalable, accurate readout of photon detection \cite{allman_near-infrared_2015} and electro-optical transduction \cite{holzgrafe_cavity_2020, rueda_efficient_2016} for technologies such as quantum sensing and quantum networking. However, absorption of higher-energy optical photons in superconducting materials causes quasiparticle formation \cite{budoyo_effects_2016, barends_minimizing_2011, mirhosseini_superconducting_2020}, resulting in losses for the optical circuit and degraded performance of the superconducting electronics. Although physically separating the optical and superconducting components as much as possible can mitigate absorption, ideally these circuits could be composed of superconducting materials engineered to be transparent to other wavelengths present on the same chip.

Transparency and conductivity, let alone superconductivity, may sound mutually exclusive. Indeed, typical superconductors like niobium are highly absorptive across the entire visible range. A film of niobium only $\SI{10}{nm}$ thick would absorb between 40-60\% of all incident light between $\SIrange{500}{1600}{nm}$, according to calculations based on measured optical constants \cite{bousquet_etude_1957, leksina_optical_nodate}. However, transparent conducting oxides such as indium tin oxide (ITO) have been widely studied due to their usefulness in photovoltaics \cite{bernardo_progress_2021}, displays \cite{betz_thin_2006}, and sensing \cite{aydin_indium_2017}. Many of these oxides are in fact degenerately doped semiconductors which, if doped to just the right amount, may superconduct while remaining transparent \cite{edwards_basic_2004}. ITO in particular has been observed to superconduct by several authors \cite{ohyama_weak_1985, mori_superconductivity_1993, chiu_four-probe_2009, aliev_reversible_2012}, typically when doped to carrier concentrations around $\SIrange{e20}{e21}{\per\cm\cubed}$. Chiu et al. \cite{chiu_four-probe_2009} even observed superconductivity in a $\SI{220}{nm}$ thick ITO nanowire, indicating that ITO could be a viable candidate for fabricating superconducting electronics.

Furthermore, several authors have observed that non-superconducting samples of ITO can become superconducting through simple post-processing methods. 
Mori \cite{mori_superconductivity_1993} found that low-temperature annealing of sputtered ITO resulted in superconducting samples, and attributed increased carrier concentration to the introduction of oxygen vacancies. 
On the other hand, Aliev et al. \cite{aliev_reversible_2012} observed superconductivity in commerically available ITO films processed by electrochemical reduction. 
They propose the increased carrier concentration is due to intercalation of small positive ions. 
However, other authors have contested whether such intercalation occurs in reduced ITO \cite{bressers_electrochromic_1998}, and Liu et al. \cite{liu_important_2015} proposed oxygen vacancies as the dopant in their own electrochemically reduced ITO. 
Additionally, while all of the above authors report seeing a visible color change in their reduced ITO films, none provide detailed quantitative data that could help determine whether the resulting films are still meaningfully transparent for engineering applications. 
If we are to optimize ITO for both superconductivity and transparency, we must understand the process by which it becomes superconducting and quantify the effect of this process on its optical constants.

\begin{figure}[ht!]
    \centering
    \includegraphics[width=\textwidth]{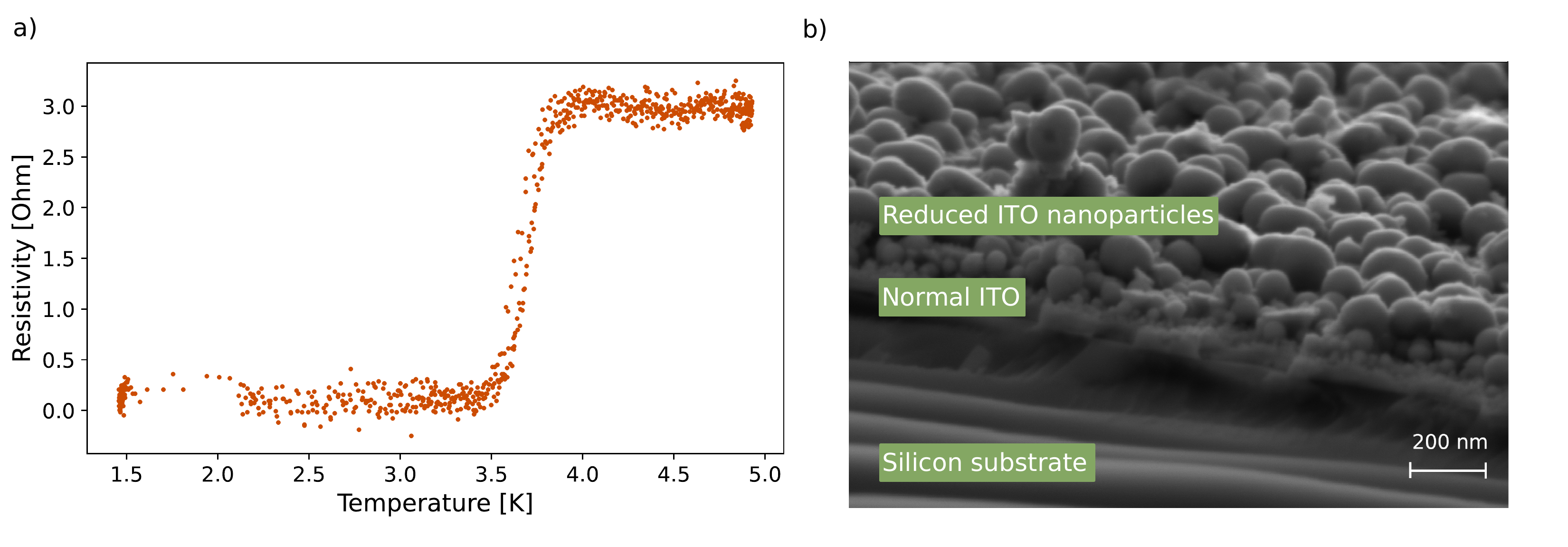}
    \caption{a) Transition temperature curve for a reduced ITO microwire $\SI{500}{\mathrm{\mu} m}$ wide. The transition is sharp, with no distinguishable hysteresis, and complete by 3.5 K. b) Superconducting properties of reduced ITO are largely confined to a layer of oxygen-deficient nanoparticles that forms on top of the original ITO surface.}
    \label{fig:summary}
\end{figure}

In this paper, we use electrochemical reduction to fabricate reduced ITO microwires and characterize changes to its electronic properties, structure and composition, and optical properties. 
The resulting microwires superconduct with a sharp transition to normal state at the switching current. 
In addition to confirming the onset of superconductivity, we found a correlation between superconductivity and the formation of a dense layer of nanoparticles, confirmed oxygen removal as the primary doping mechanism in our material, and quantified the resulting changes to refractive index and overall absorption of the reduced films. The transition temperature curve from a $\SI{500}{\mathrm{\mu} m}$ microwires is shown in Figure \ref{fig:summary} (a), while an SEM of the nanoparticles is shown in Figure \ref{fig:summary} (b).

\section{Methods}

Here we describe our methods for generating reduced ITO material, fabricating electronics from reduced ITO, and characterizing the electronic and optical properties and structure and composition of reduced ITO \cite{batson_reduced_2022}.

\subsection{Material Processing}

We used ITO from two sources for these experiments: material sputtered at MIT, which we refer to as ITO on silicon, and commercially available material from Sigma Aldrich, which we refer to as ITO on glass. For ITO on silicon, RF-magnetron sputtering was used to deposit ITO on silicon substrates at a rate of 0.6 Å/s with an argon pressure of 6 mTorr and 60W RF power to a thickness of 130 nm. There were two wafers deposited this way, and the ITO properties differed between the two. The ITO on glass \cite{noauthor_indium_nodate} was 120-160 nm thick. The ITO on silicon was used for measurements where having a conductive or opaque substrate was useful, such as electron microscopy or ellipsometry. The ITO on glass was generally more uniform and was used to test repeatability of the process or for testing sensitivity to process parameters. Qualitatively, reduction had similar effects on material from both sources.

We reduced the ITO in a three-electrode galvanostatic cell with a 1 M sodium chloride electrolyte, a schematic diagram of which can be found in Appendix A. We drove $\SI{300}{\mathrm{\mu} A}$ of current from a platinum counter-electrode into the ITO working electrode, and measured the resulting overpotential between the ITO working electrode and an ITO reference electrode which was roughly in equilibrium with the electrolyte. Parts of the ITO submerged in electrolyte exhibited a visible color change, the area of which was measured with calipers and treated as the total reduced area. This area, which was typically around $\SI{0.5}{\square\cm}$ for unpatterned films, was used to calculate current density and reduction charge density. Current density was fixed at about $\SI{0.6}{mA\per\square\cm}$, slightly higher than the $\SI{0.1}{mA\per\square\cm}$ used by Aliev et al. \cite{aliev_reversible_2012}, and total reduction charge density was varied across samples by varying the reduction time. Color change ranged from translucent reddish-brown at low reduction levels ( $< \SI{100}{mC\per\square\cm}$), reflective reddish-purple at intermediate reduction levels ( $\approx \SI{100}{mC\per\square\cm}$), and matte gray at high reduction levels ( $>\SI{1000}{mC\per\square\cm}$). Finally, reduced samples were annealed for an hour at $\SI{160}{\degree C}$ in air. For this paper, we focus on samples reduced to around $\SI{100}{mC\per\square\cm}$ or lower.

\subsection{Electronics Fabrication}

We fabricated our reduced ITO wires by directly patterning the reduction process. We spun an insulating photoresist onto our samples and patterned a mask to block reduction current. We reduced the masked samples, and only ITO exposed by the mask became superconducting. More details are provided in Appendix B.

We used this direct patterning technique to create five ITO wires of different widths on a single chip on the ITO on glass. All wires were $\SI{2750}{\mathrm{\mu}\m}$ long with four equally spaced metallic contact pads. The reduction time and current were adjusted for a target reduction charge density of about $\SI{75}{mC\per\square\cm}$ normalized to the exposed area, which preliminary experiments indicated would maximize transition temperature. Scanning electron microscopy and optical microscopy indicated that the widest wires formed dense nanoparticle films in keeping with the expected reduction charge density, but the narrowest wires, particularly the $\SI{100}{\mathrm{\mu}\m}$ wide wire, appeared under-reduced. An SEM of the $\SI{100}{\mathrm{\mu}\m}$ wide wire is shown in Figure \ref{fig:iv}(a).

\subsection{Material Characterization}

Superconductivity of the samples was confirmed by measuring sample resistivity in a 4-point probe configuration while sweeping through a temperature range from about 2 - 6 K while applying a current of $\SI{10}{\mathrm{\mu} A}$. $I-V$ curves of wires were taken through a 4-point probe configuration at a fixed temperature, 300 mK, while sweeping current. Additionally, normal state resistivity was determined by a room temperature 4-point probe measurement.

We studied the surface morphology of the samples with scanning electron microscopy (SEM) and atomic force microscopy (AFM) imaging. We also took cross-sectional images of cleaved samples with SEM. We characterized the composition of a lamella, prepared by focused ion beam milling, with scanning transmission electron microscopy and energy-dispersive x-ray spectroscopy (STEM-EDS). These methods were most successful with ITO on silicon, as an insulating glass substrate can lead to charge buildup. 

To further study how the oxidation state changed over the depth of the film, we also performed x-ray photoelectron spectroscopy (XPS) depth profiles on select samples with a Physical Electronics Versaprobe II by alternating XPS collection with argon ion milling. Glass substrates could be used here because the XPS includes charge neutralization beams.

Ellipsometric measurements and optical modeling were performed with the Semilab SE-2000 spectroscopic ellipsometer at $60\degree$, $65\degree$, and $70\degree$ angles of incidence from $\SIrange{240}{1700}{nm}$ wavelengths of light. All three angles were fit simultaneously to a Drude-Lorentz model, which has been demonstrated by Baum et al. \cite{baum_determination_2013} to be effective for modeling ITO nanoparticles. Ellipsometric measurements were only performed on ITO on silicon, as the transparent glass substrate proved difficult to work with. Details of the ellipsometric models used are given in the Optics Supplementary Materials section.

\section{Results and Discussion}

We next report our findings on the changes to electronic properties, surface properties, film composition, and optical properties caused by electrochemical reduction of ITO.

\subsection{Electronic Properties}

We observed different effects of reduction on sheet resistance depending on our starting material. For ITO on silicon, sheet resistance universally decreased with reduction from around $\SI{50}{\ohm/\sq}$ to around $\SI{40}{\ohm/\sq}$ regardless of reduction charge density. For ITO on glass, sheet resistance increased with reduction charge density from around $\SI{10}{\ohm/\sq}$ to anywhere between $\SIrange{12}{90}{\ohm/\sq}$, with larger reduction charge density correlating to larger final sheet resistance. It is unclear why the two types of films perform differently. However, given the different initial sheet resistances, it stands to reason that the ITO on glass was optimized for minimum resistance, and that any conductivity gained by the carrier concentration increase was more than offset by mobility reduction from the doping process. Sheet resistance values are given in Appendix C.

We tested films from both sources and with a range of reduction charge densities for superconductivity. As-received films were not superconducting, but all the electrochemically reduced films we tested, ranging from $\SIrange{16}{110}{mC\per\square\cm}$, exhibited superconducting transitions. Our results, shown in Figure \ref{fig:summary} (a), were consistent with a dome-shaped dependence of transition temperature on reduction charge density as reported by others in ITO \cite{aliev_reversible_2012} and in line with the behavior of several other superconducting doped semiconductors \cite{koonce_superconducting_1967, ueno_discovery_2011, das_superconducting_2015, yu_superconducting_2020}. The maximum transition temperature observed was 4 K in a sample reduced to about $\SI{60}{mC\per\square\cm}$. Further transition temperature values are given in Appendix C.

\begin{figure}[ht!]
    \centering
    \includegraphics[width=\textwidth]{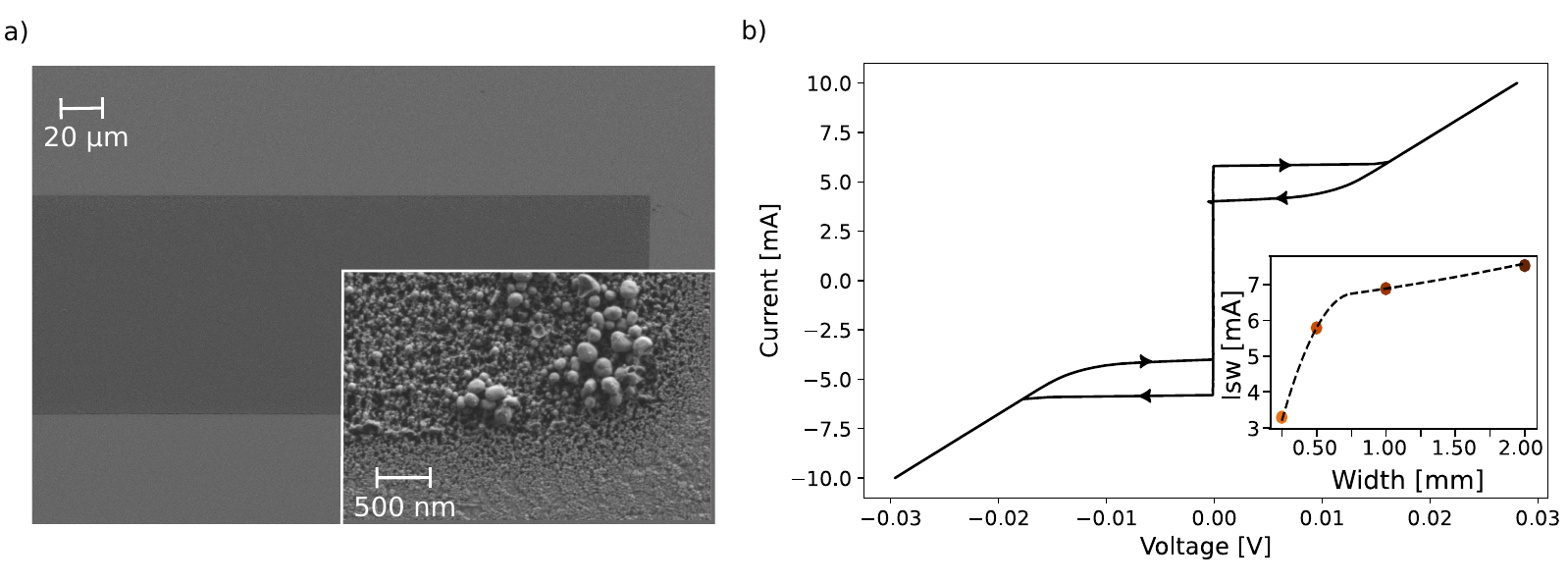}
    \caption{a) SEM of a $\SI{100}{\mathrm{\mu} m}$ wide wire patterned by electrochemical reduction. The inset shows the nanoparticle morphology and transition to non-reduced ITO at the corner. Note the increase in size of nanoparticles near the corner, possibly indicating geometry-dependent reduction current dynamics. b) $I-V$ curve for 500 $\si{\mathrm{\mu} m}$ wide wire. The three widest wires (500, 1000, and 2000${\mathrm{\mu} m}$) have similar $I-V$ curves, exhibiting a clear and sharp transition at the switching current into the resistive regime and hysteretic cooling to return to the superconducting state. (Inset) Extracted switching currents for all wires that showed superconducting $I-V$ curves. Switching current increases somewhat quadratically with width, suggesting a nonuniform current distribution as typical of wires wider than the Pearl length.}
    \label{fig:iv}
\end{figure}

Additionally, we measured the $I-V$ curves and determined the transition temperatures for the five directly patterned ITO wires. As demonstrated by the $I-V$ curve for the $\SI{500}{\mathrm{\mu} m}$ wide wire shown in Figure \ref{fig:iv}(b), we were able to obtain sharp transitions to the normal state while ramping up current resulting in hysteretic $I-V$ curves for the widest wires, despite being shunted to about $\SI{2}{\ohm}$ by the underlying normal ITO layer. The $I-V$ curves for the 500, 1000, and 2000 ${\mathrm{\mu} m}$ wide wires all had similar shapes. The $\SI{250}{\mathrm{\mu} m}$ wide wire lacked hysteresis, but still clearly switched. All four of these wires also had nearly identical transition temperature curves to that of the $\SI{500}{\mathrm{\mu} m}$ wide wire shown in Figure \ref{fig:summary}(a). The curves exhibited a transition region around 0.5 K wide, a transition temperature defined by reaching 10\% of the 5 K resistivity of 3.5 K, and a small hysteresis. The $\SI{100}{\mathrm{\mu} m}$ wide wire did not exhibit a clear superconducting regime on its $I-V$ curve and had a broader transition, about 2 K wide, and seemed to just barely have completed the transition to zero resistance at 1.3 K, the bottom of our measurement region. The extracted switching currents, shown in the upper inset of Figure \ref{fig:iv}(b), do not increase linearly with width, which is consistent for wires wider than the Pearl length due to increased current screening \cite{pearl_vortex_nodate}. Additional $I-V$ curves and transition temperature curves are provided in Appendix D.

Overall, the four widest wires exhibited $I-V$ curves and transition temperatures suitable for nanowire electronics, while the $\SI{100}{\mathrm{\mu} m}$ wide wire performed less well. Kowal et al. found that disorder can limit the aspect ratio of wires made of superconducting indium oxide, the parent material of ITO \cite{kowal_scale_2008}. However, our $\SI{100}{\mathrm{\mu} m}$ wide wire was also a visibly lighter color and had a degraded transition temperature compared to the other wires, suggesting another explanation for the degraded superconductivity in the $\SI{100}{\mathrm{\mu} m}$ wide wire. Probably our device widths were limited by reduction current dynamics favoring larger features and under-reducing smaller features, rather than a fundamental scaling limitation for superconducting ITO. Narrower superconducting wires might be successfully fabricated if no wide wires are fabricated on the same chip. 

\subsection{Surface and Composition}

In addition to the expected electronic changes, we also observed significant surface effects of electrochemical reduction, as seen in Figure \ref{fig:summary}(a) and Figure \ref{fig:iv}(a). In particular, we found that reduction caused the formation of a dense nanoparticle thin film on top of the original surface, which has been reported elsewhere \cite{matveeva_electrochemistry_2005, spada_role_2013, liu_important_2015, bouden_multifunctional_2016}, but not previously noted to correspond to superconductivity. However, our depth-dependent composition and optical measurements reported later in the paper found that nearly all reduction-induced changes were localized to the surface nanoparticle region.

\begin{figure}[ht!]
    \centering
    \includegraphics[width=0.6\textwidth]{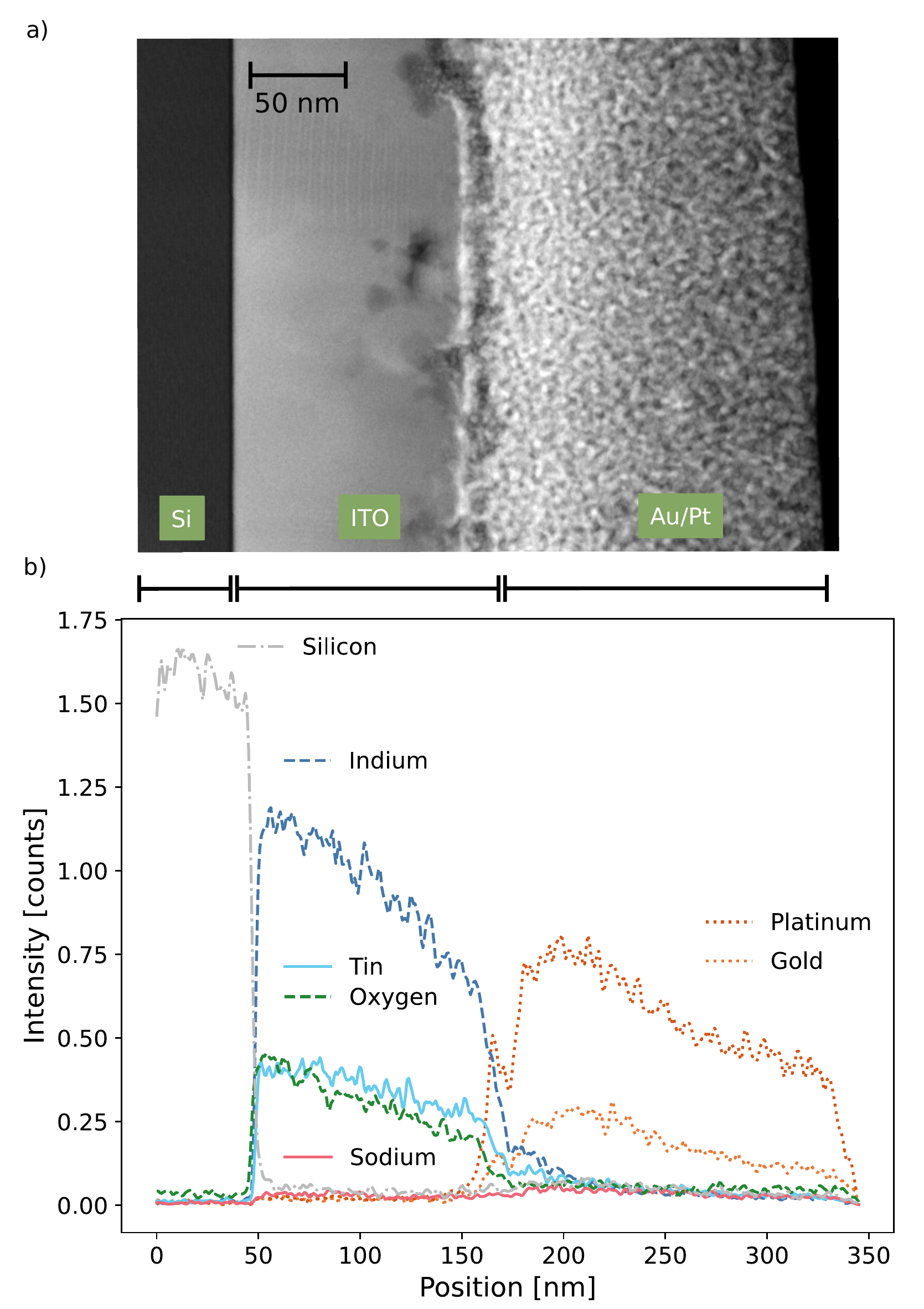}
    \caption{(a) High-angle annular dark field (HAADF) STEM image of an ITO lamella reduced for 120 s to about $\SI{75}{mC\per\square\cm}$ imaged with STEM and EDS. Brighter regions are those that interacted more strongly with electrons. The main film appears polycrystalline, but also exhibits odd nanotextured voids, perhaps where ITO was removed by reduction to reform as nanoparticles, which appear to be hollow. (b) EDS intensity counts corresponding to this area. Sodium signal does not rise above noise levels anywhere in the film, indicating that intercalation may not be a significant source of doping.}
    \label{fig:stem}
\end{figure}

We also investigated the composition of the reduced ITO films, particularly focusing on the nanoparticles. We especially hoped to determine whether sodium ion intercalation \cite{aliev_reversible_2012} or oxygen vacancy defects \cite{liu_important_2015} were the more significant doping mechanism for our films. Cross-sectional STEM with EDS, as shown in Figure \ref{fig:stem}(b), did not find sodium above noise levels throughout the film. However, STEM did indicate a general decrease in density of the film toward the surface and possible metallization of hollow nanoparticles, as seen in Figure \ref{fig:stem}(a). Near the interface with silicon, the ITO appears polycrystalline and fairly uniform. Towards the gold and platinum cap that was applied for processing, there are very low-density voids distributed throughout the ITO. Past the thickness of the original surface, we observe a hollow strip flanked on either side by narrow bright regions of mostly indium and tin. The location of this hollow, metallized structure corresponds to the nanoparticles.

\begin{figure}[ht!]
    \centering
    \includegraphics[width=\textwidth]{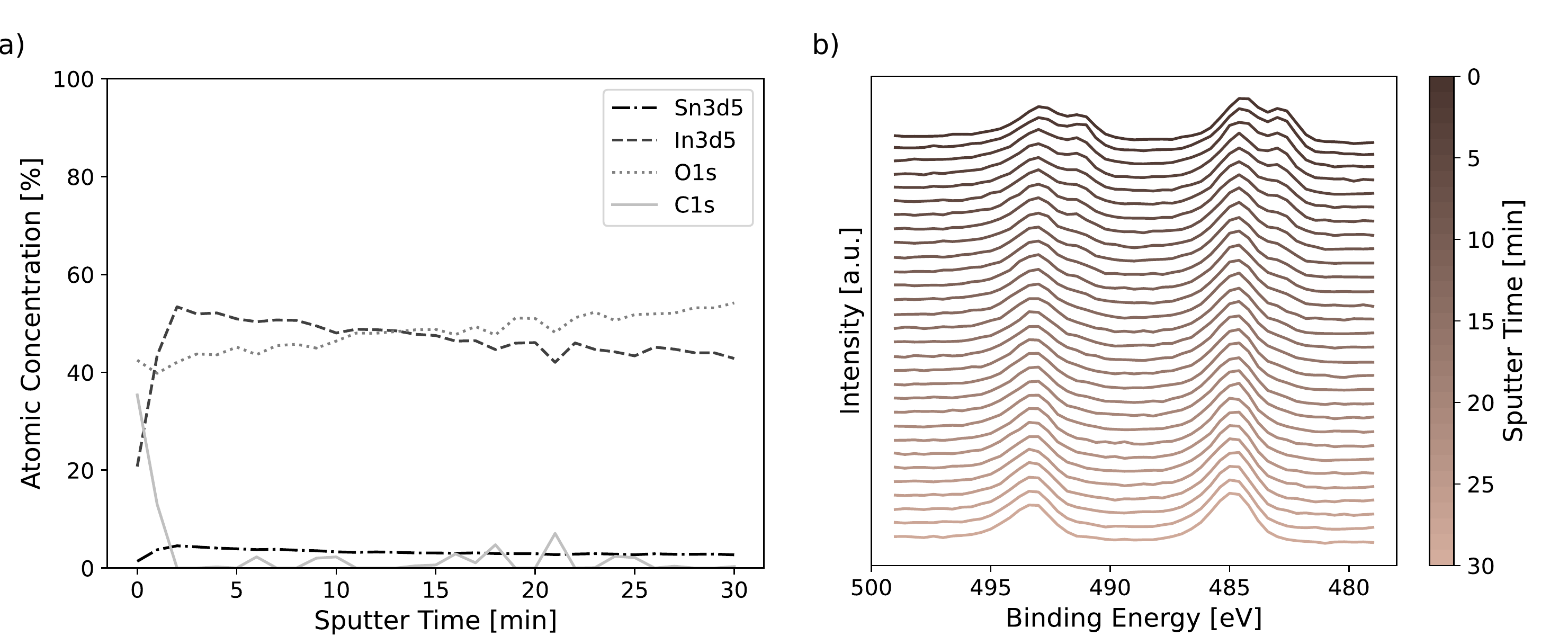}
    \caption{(a) XPS depth profile with high-resolution peaks for tin in ITO reduced for 120 s. The film was sputtered for 1 minute with an argon ion beam between measurements. Carbon contamination is shown for reference; note that the first data point at the very surface is strongly contaminated. After this, overall atomic concentration calculations indicate a gradient, with a lower percent concentration of oxygen near the surface of the material, further suggesting oxygen removal. However, at no point does the film appear fully metallized. XPS showed no clear indication of sodium in this film. (b) High-resolution XPS tin spectra. Near the surface, the tin peaks widen toward lower binding energies, suggesting the presence of lower, more metallic oxidation states.}
    \label{fig:xps}
\end{figure}

Given the low sodium concentrations and metallic appearance of the nanoparticles, we turned to XPS to better quantify film oxygen content. Due to its high energy resolution, XPS can determine atomic concentration of different atoms and even distinguish between oxidation states of metals. We performed depth profiles on two samples on glass: one as-received (Appendix E), the other reduced for 120 s (Figure \ref{fig:xps}). The as-received sample appeared to be nearly stoichiometric indium oxide with a small percentage of tin, with no significant depth dependence of the peak shapes or atomic concentrations, except for the usual carbon contamination on the very surface. However, the properties of the reduced sample were strongly depth dependent. Ignoring the topmost surface, which shows strong hydrocarbon contamination, in Figure \ref{fig:xps}(a), we can observe a lower fraction of oxygen near the surface of the reduced ITO, recovering to stoichiometric values by the end of the depth profile. As shown in Figure \ref{fig:xps}(b), the tin peaks are also much broader near the surface, indicating the presence of multiple oxidation states. Full profiles of oxygen and indium are given in Appendix E. In particular, indium and tin broaden toward lower energies near the top of the film, which are associated with more metallic states. However, there is evidence of oxygen and oxidized indium and tin throughout the depth of the film.

We also attempted to analyze the XPS data for sodium. Unfortunately, the sodium 1s peak significantly overlaps with a broad, intense peak from indium Auger electron energies, so XPS cannot be used as a clear indicator of sodium presence. Samples reduced to over $\SI{1000}{mC\per\square\cm}$ do have a small but definite sodium peak visible despite the indium background, but samples at reduction levels of interest to superconducting electronics do not exhibit an obvious sodium peak.

Finally, we confirmed Aliev et al.'s \cite{aliev_reversible_2012} observation that applying an oxidizing electrochemical current can almost fully reverse color change, which they attributed to removal of sodium ions, but one cannot rule out re-oxidation of the material in a water bath. However, we also found that annealing the samples in air at a higher temperature of around $\SI{250}{\degree C}$ reversed the color change in a similar manner. Annealing could have replenished oxygen to the film, but seems unlikely to have removed metal ions. This provides further evidence for oxygen deficiency as the primary driver of changes in our film.

% Peer reviewer suggested oxygen vacancies wouldn't be reversible in this way, but I'm not sure that's true so unsure how to address

\subsection{Optical Properties}

Fitting with least-squares regression to the Drude-Lorentz-Lorentz model, we achieved $R^2 > 0.997$ for all the reduced ITO samples and extracted both physically meaningful parameter values and the complex refractive index resulting from the fit. The parameter values used in the fit are given in Appendix F. Generally, as reduction increases, the nanoparticle layer becomes thicker. Also, the near-UV Lorentz oscillator becomes lower in center energy and higher in amplitude, likely corresponding to oxygen vacancy dopants lowering the band gap energy \cite{zhang_metal_2020}. The Drude plasma frequency and core permittivity of the nanoparticle layer are mostly independent of reduction time, but are starkly different for the nanoparticles than for pristine ITO. Therefore, these changes are most likely related to the nanoparticle morphology rather than electronic properties, and could be evidence of surface plasmon resonance \cite{kanehara_indium_2009}.

\begin{figure}[ht!]
    \centering
    \includegraphics[width=\textwidth]{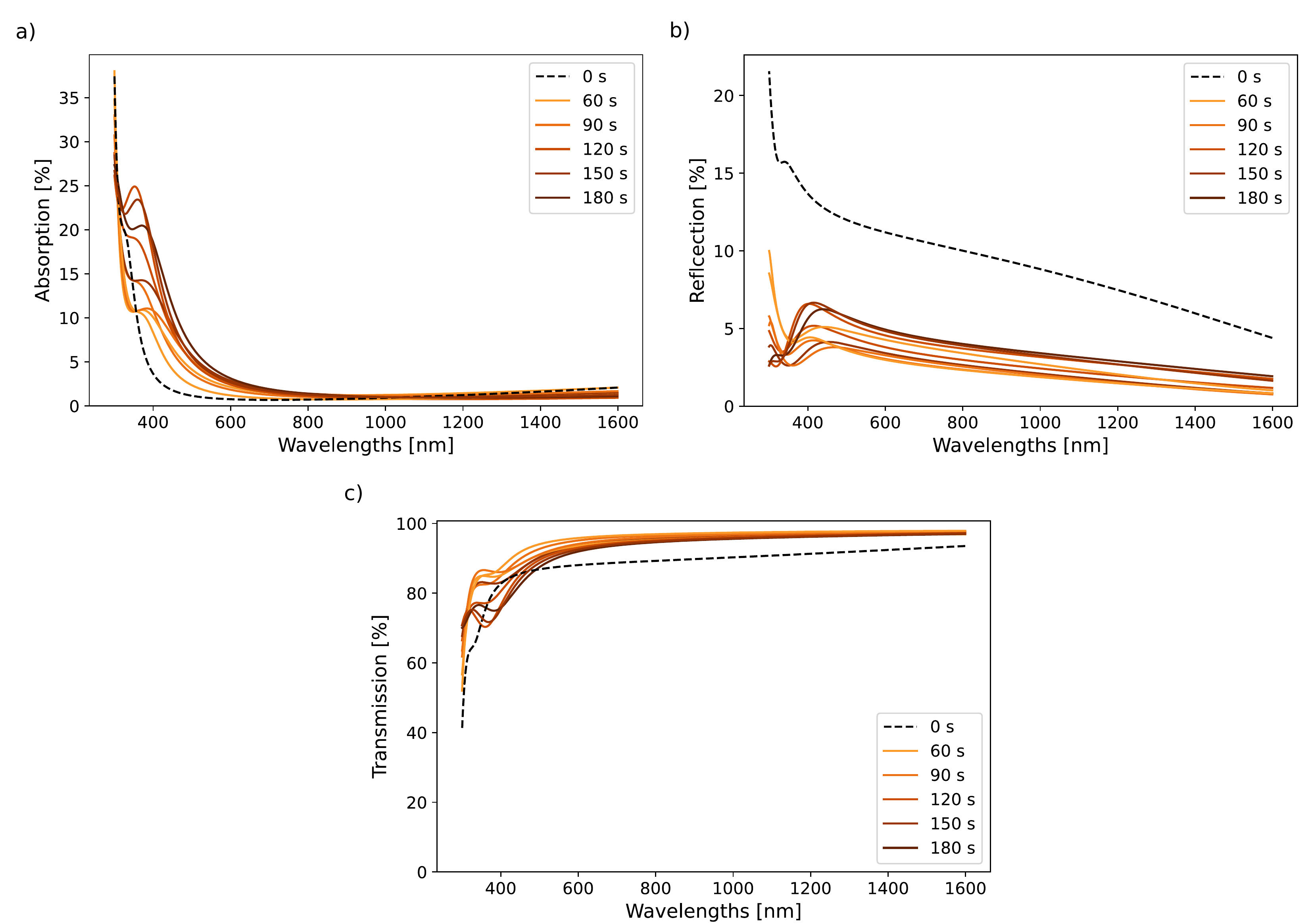}
    \caption{a) Absorption, b) reflection, and c) transmission calculated for a 10 nm thick film of ITO or reduced ITO based on the complex refractive index measured with ellipsometry. The legend shows how long each sample was reduced with $\SI{300}{\mathrm{\mu} A}$. Reduced films are more absorbing and less reflective than the original film, but still have similar optical characteristics.}
    \label{fig:optical}
\end{figure}

From the refractive index, we computed the predicted transmission, absorption, and reflection for reduced ITO nanoparticle films with varying thicknesses and angles of incidence, based on the equations derived by Bousquet \cite{bousquet_etude_1957}. The results for a 10-nm-thick film with normal incidence are shown in Figure \ref{fig:optical}. Although these plots confirm the increased absorption in blue wavelengths, reduced ITO is overall optically very similar to the original film. A 10-nm-thick film of niobium would absorb 40-60\% of light over the 500 - 1700 nm range we measured with ellipsometry \cite{leksina_optical_nodate}, compared to absorption ranging from less than 1\% to about 20\% for reduced ITO. In fact, for wavelengths between 800 - 1700 nm, absorption in reduced ITO is nearly equal to or even less than that of pristine ITO.

\section{Conclusion}

We were able to fabricate transparent superconducting wires by patterning the electrochemical reduction of ITO. Although these wires were much wider than typical superconducting electronics, they exhibited sharp switching from superconducting to normal state at the critical current and transition temperatures from 3 - 4 K. We confirmed that thin films of reduced ITO is twenty to forty times less absorptive than niobium between 600 - 1600 nm, showing quantitatively that superconductivity and optical transparency can be simultaneously achieved.

While the simplicity and convenience of electrochemically reduced ITO as a transparent superconductor is appealing, the rough nanoparticle structure means it is likely unsuited for typical nanowire electronics. However, we have determined that the primary reduction mechanism for ITO in our process is oxygen removal rather than ion intercalation. Therefore, we should expect that ITO that is deposited with low oxygen content should exhibit similar superconductivity while achieving a more uniform surface. This is likely the mechanism behind other instances of superconducting ITO \cite{mori_superconductivity_1993, chiu_four-probe_2009}, though those materials have not been as thoroughly characterized as we have done here.

There are also other materials that show promise as transparent superconductors. Our work with ITO indicates that doped semiconductor-type superconductors may not be the best suited for electronics work unless the doping is performed carefully so as not to disrupt the thin film structure too much. Materials which are transparent superconductors in their stoichiometric forms, such as lithium titanium oxide (LTO) \cite{ohsawa_origin_2020}, may prove better for electronics work.

\ack

We gratefully acknowledge provision of epitaxial ITO samples from Julia Mundy, Johanna Nordlander, Margaret Anderson, and Erika Ortega Ortiz from Harvard University, as well as helpful discussion of spin-coated ITO with Stephanie Hurst and Madison King from Northern Arizona University. We also thank Andrew Dane for initiating work on this project. We acknowledge Yang Yu's help in preparing the STEM lamella and Aubrey Penn's assistance obtaining the STEM data, Luqiao Liu and Brooke McGoldrick for training and access to the ion milling tool, Jim Daley for gold evaporation, and Libby Shaw for XPS training. We thank the staff of the NSL, MRL, and MIT.nano for instrument and facility support, and Matteo Castellani and Shruti Nirantar for their valuable writing feedback.

Initial work was supported in part by the NSF as part of the RAISE-TAQS program under Grant No. ECCS1839197, and later work by the NSF as part of the CQN program under Grant No. EEC1941583. Emma Batson gratefully acknowledges the National Science Foundation Graduate Research Fellowship under Grants No. 1745302 and 2141064.

\appendix 

\section{}

\begin{figure}[ht!]
    \centering
    \includegraphics[width=0.5\textwidth]{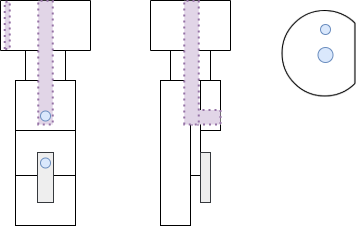}
    \caption{Schematic of sample holders used to hold working and reference ITO samples in place in the electrochemical cell.}
    \label{fig:apparatus_sample}
\end{figure}

\begin{figure}[ht!]
    \centering
    \includegraphics[width=\textwidth]{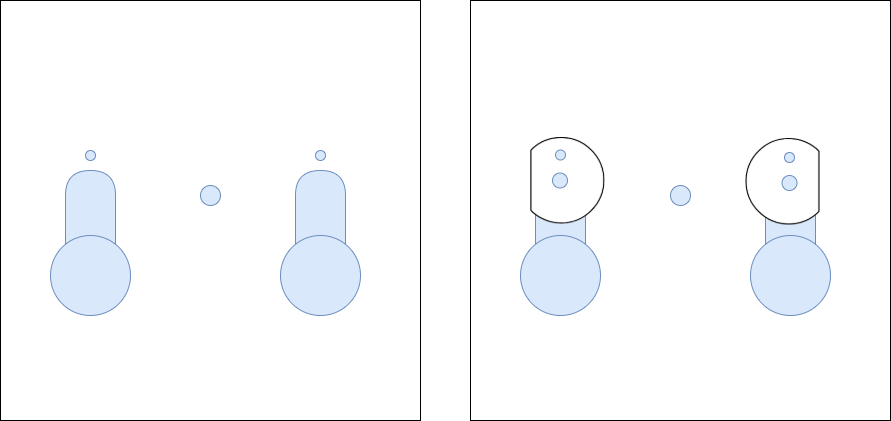}
    \caption{Top view of electrochemical cell aparatus used to keep samples at a consistent angle and separation across experiments. The right-hand figure shows where the sample holders sit when the cell is assembled.}
    \label{fig:apparatus_top}
\end{figure}

\begin{figure}[ht!]
    \centering
    \includegraphics[width=\textwidth]{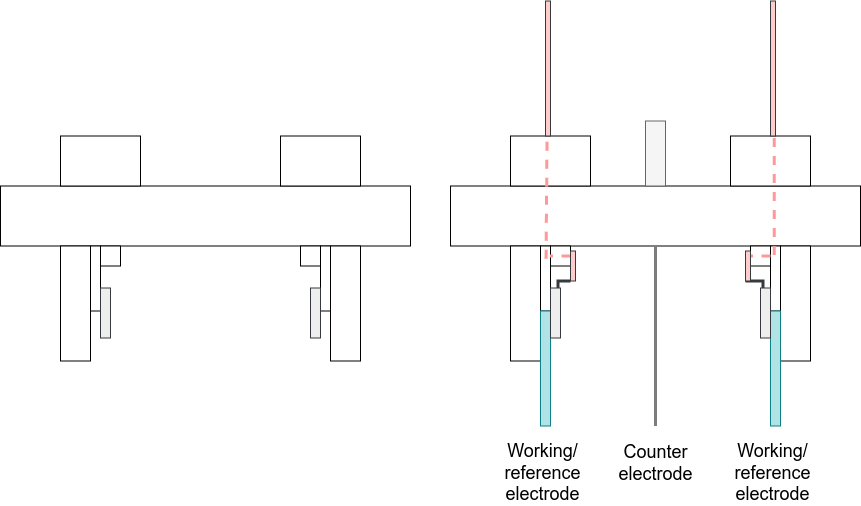}
    \caption{Side view of electrochemical cell aparatus used to keep samples at a consistent angle and separation across experiments. The right-hand figure shows the cell fully assembled, with wires and electrodes in place.}
    \label{fig:apparatus_side}
\end{figure}

Our electrochemical cell apparatus was made with simple machining techniques out of teflon, although any insulating, machinable material should be suitable for similar apparatus. It consisted of three separable components: two sample holders and one ``beaker lid" to rest on the top of the beaker and fix the sample holders in place. Front, side, and top views of the sample holders are shown in Figure \ref{fig:apparatus_sample}. A top view of the beaker lid with and without the sample holders inserted is shown in Figure \ref{fig:apparatus_top}, and a side view is shown in Figure \ref{fig:apparatus_side}.

\section{}

In order to directly pattern the reduction, we spun on a layer of S1813 photoresist and baked it for 90 s at $\SI{100}{\degree}$C. This resist was patterned into wires in a Heidelberg $\si{\mathrm{\mu} PG}$ 101 at 8 mW with 25\% duty cycle, and developed in CD26 for 60 s. The masked sample was reduced in our three-electrode electrochemical cell, with current and time adjusted to maintain proper reduction charge density for the smaller exposed area. After reduction, the remaining photoresist was cleaned off with acetone and IPA. Next we spun on a layer of PMGI SF8 and baked it for 90 s at $\SI{180}{\degree C}$ followed by a layer of S1813 baked for 90 s at $\SI{100}{\degree C}$. We then patterned the pads in a four-point probe configuration on each wire in the Heidelberg and developed in CD26 for 80 s. Then $\SI{10}{nm}$ of titanium and $\SI{50}{nm}$ gold were thermally evaporated onto the chip, and the pads were formed by liftoff. Finally samples were attached with thermal varnish and wirebonded to a PCB for testing.

\section{}

\begin{table}[ht!]
\begin{center}
\caption[ITO on Silicon Sheet Resistances]{ITO on Silicon Sheet Resistances}
Sheet resistances before and after reduction for ITO sputtered onto silicon. Final resistance is basically independent of reduction charge density.
    \begin{tabular}{|c|c|c|c|}
        \hline
         \thead{Reduction Charge \\ Density [mC/cm$^2$]} & Original $R_s [\si{\ohm/\sq}]$  & New $R_s [\si{\ohm/\sq}]$ & Percent Change in $R_s [\si{\ohm/\sq}]$ \\
         \hline
         36 (a) & 50 & 43 & -13\% \\
         36 (b) & 50 & 44 & -12\% \\
         54 (a) & 58 & 40 & -30\% \\
         54 (b) & 58 & 38 & -34\% \\
         72 (a) & 42 & 38 & -8\% \\
         72 (b) & 42 & 40 & -4\% \\
         90 (a) & 47 & 39 & -17\% \\
         90 (b) & 47 & 46 & -1\% \\
         108 (a) & 46 & 41 & -11\% \\
         108 (b) & 46 & 41 & -11\% \\
         \hline
    \end{tabular}
\end{center}
\label{tab:itosi_res}
\end{table}

\begin{table}[ht!]
\caption[ITO on Glass Sheet Resistance]{ITO on Glass Sheet Resistances}
Sheet resistance before and after reduction for ITO on glass from Sigma Aldrich. Increasing reduction charge density correlates with an increase in sheet resistance.
\begin{center}
    \begin{tabular}{|c|c|c|c|}
        \hline
         \thead{Reduction Charge \\ Density [mC/cm$^2$]} & Original $R_s [\si{\ohm/\sq}]$  & New $R_s [\si{\ohm/\sq}]$ & Percent Change in $R_s [\si{\ohm/\sq}]$ \\
         \hline
         16 & 11 & 12 & 9\% \\
         19 & 11 & 12 & 9\% \\
         30 & 9 & 13 & 48\% \\
         41 & 9 & 14 & 54\% \\
         54 & 12 & 17 & 38\% \\
         55 & 12 & 17 & 40\% \\
         63 & 11 & 15 & 41\% \\
         80 & 11 & 19 & 77\% \\
         91 & 12 & 91 & 317\% \\
         130 & 12 & 32 & 162\% \\
         \hline
    \end{tabular}
\end{center}
\label{tab:itosa_res}
\end{table}

\begin{table}[ht!]
\caption{Transition temperature measurements of films from different sources reduced to different degrees.}
\begin{center}
    \begin{tabular}{ccc}
        \hline
         Substrate & Reduction Charge Density [$\si{mC\per\square\cm}$] & $T_c$ [K] \\
        \hline
        Silicon (a) & 36 & 1.37 \\
        Silicon (a) & 54 & 1.95 \\
        Silicon (a) & 72 & 1.38 \\
        Silicon (a) & 72 & 1.42 \\
        Silicon (a) & 90 & 1.85 \\
        Silicon (a) & 90 & 1.38 \\
        Silicon (a) & 108 & 1.56 \\
        Silicon (b) & 90 & 3.08 \\
        Silicon (b) & 90 & 3.11 \\
        Glass & 38 & 3.34 \\
        Glass & 43 & 3.31 \\
        Glass & 50 & 3.84 \\
        Glass & 59 & 3.99 \\
        Glass & 99 & 3.78 \\
        \hline
    \end{tabular}
\end{center}
\label{tab:tc}
\end{table}

\section{}

Here we present additional data on the $I-V$ curves of microwires of different widths and their transition temperatures.

\begin{figure}[ht!]
    \centering
    \begin{subfigure}[b]{0.4\textwidth}
        \includegraphics[width=\textwidth]{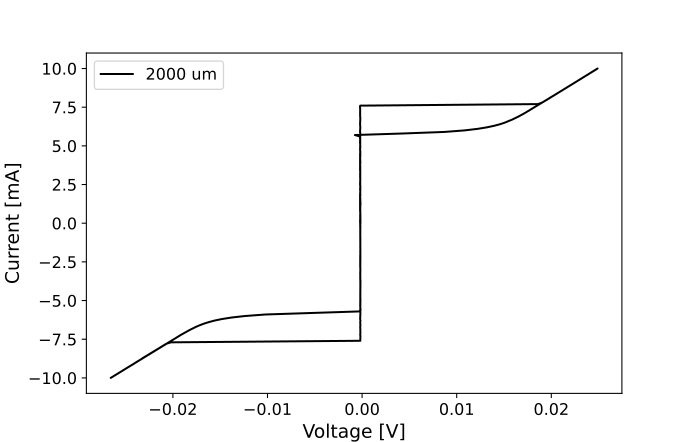}
        \caption{$\SI{2000}{\mathrm{\mu} m}$}
    \end{subfigure}
    \begin{subfigure}[b]{0.4\textwidth}
        \includegraphics[width=\textwidth]{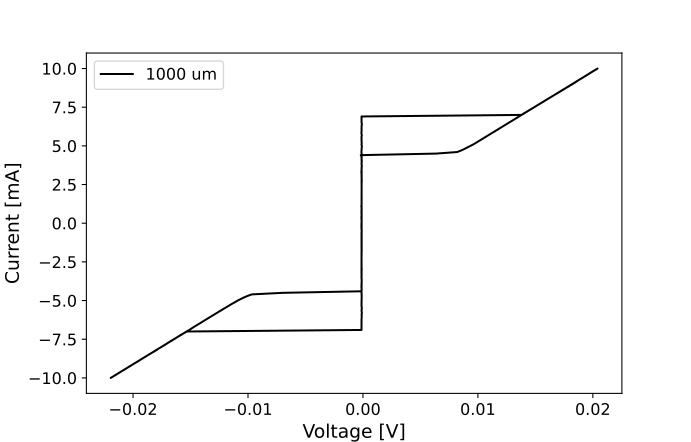}
        \caption{$\SI{1000}{\mathrm{\mu} m}$}
    \end{subfigure}
    \begin{subfigure}[b]{0.4\textwidth}
        \includegraphics[width=\textwidth]{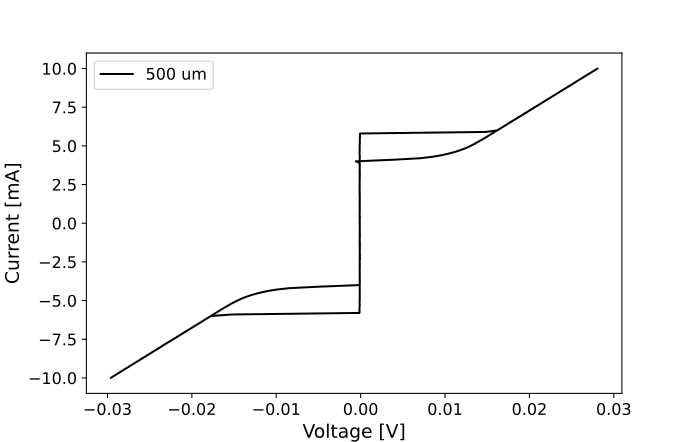}
        \caption{$\SI{500}{\mathrm{\mu} m}$}
    \end{subfigure}
    \begin{subfigure}[b]{0.4\textwidth}
        \includegraphics[width=\textwidth]{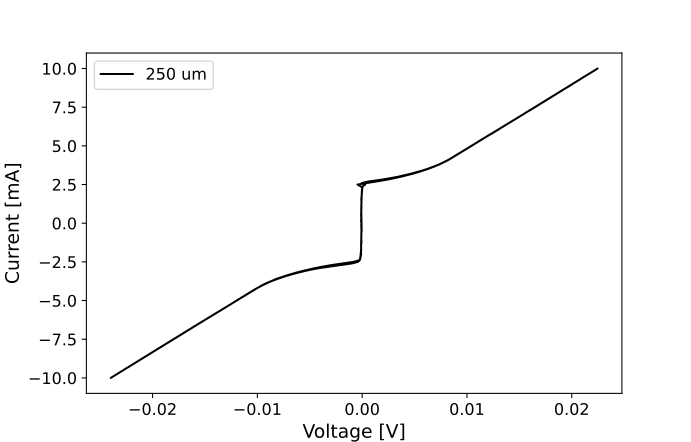}
        \caption{$\SI{250}{\mathrm{\mu} m}$}
    \end{subfigure}
    \begin{subfigure}[b]{0.4\textwidth}
        \includegraphics[width=\textwidth]{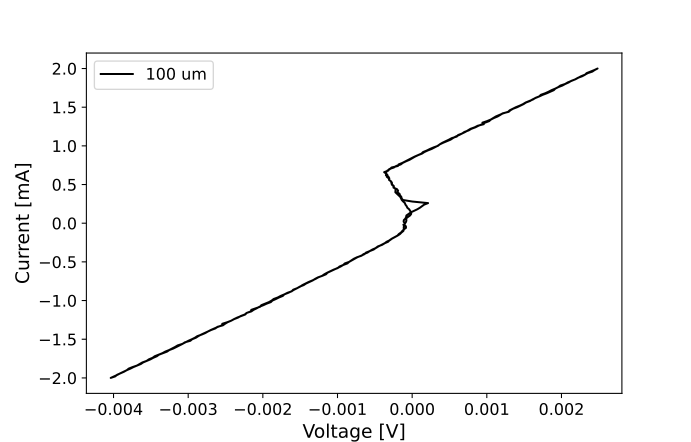}
        \caption{$\SI{100}{\mathrm{\mu} m}$}
    \end{subfigure}
    \begin{subfigure}[b]{0.4\textwidth}
        \includegraphics[width=\textwidth]{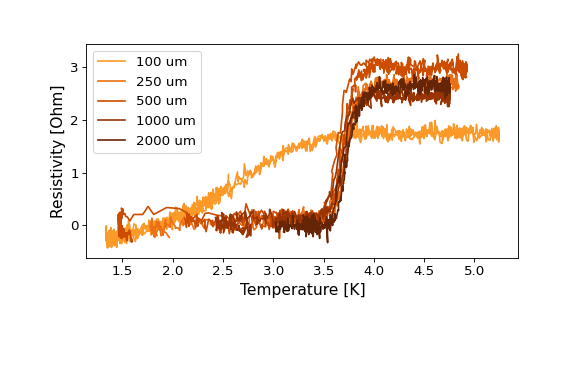}
        \caption{Transition temperatures of all wires}
    \end{subfigure}
    \caption{(a-e) $I-V$ curves for the directly patterned microwires. (f) Transition temperature curves for the directly patterned microwires. They are all roughly the same, except for that of the $\SI{100}{\mathrm{\mu} m}$ wide wire, which showed a wider, lower transition and also did not have a clear superconducting regime in its $I-V$ curve. This wire was also visibly lighter in color than the others, so we attribute its poor performance to low reduction levels due to reduction current dynamics rather than a fundamental limitation on the width of reduced ITO wires. Notably the $\SI{250}{micro m}$ wide wire has the same transition temperature as the others, and its $I-V$ curve has a clear superconducting regime, but it lacks the sharp hysteretic transition of the wider wires.}
\end{figure}

\section{}

Here we present complete data from the XPS depth profile of reduced ITO, showing how the indium, tin, and oxygen peaks all varied with sputtering time.

\begin{figure}[ht!]
    \centering
    \begin{subfigure}[b]{0.4\textwidth}
        \includegraphics[width=\textwidth]{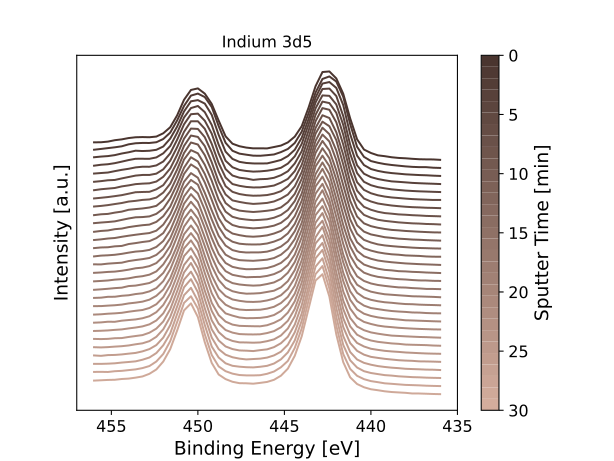}
        \caption{}
    \end{subfigure}
    \begin{subfigure}[b]{0.4\textwidth}
        \includegraphics[width=\textwidth]{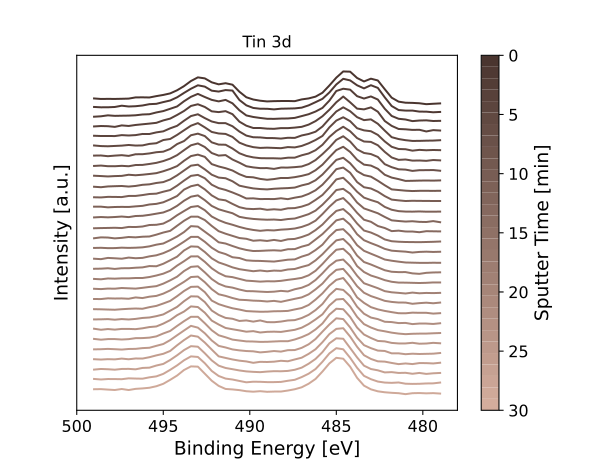}
        \caption{}
    \end{subfigure}
    \begin{subfigure}[b]{0.4\textwidth}
        \includegraphics[width=\textwidth]{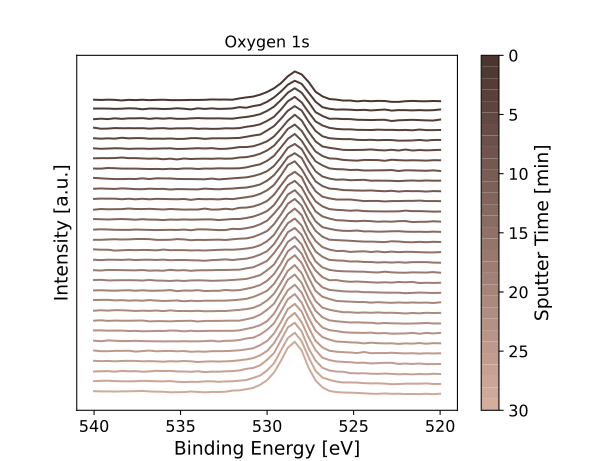}
        \caption{}
    \end{subfigure}
    \caption{XPS spectra taken throughout the depth of ITO reduced for 120 s for (a) indium, (b) tin, and (c) oxygen. These spectra are all consistent with an increase in prevalence of more metallic states of indium and tin, but not total vanishing of oxides, near the surface of the reduced film.}
    \label{fig:my_label}
\end{figure}

\section{}

Before reduction, we modeled the ITO optical constants with a superposition of a Drude oscillator and two Lorentz oscillators. Initial parameter guesses were provided by the ellipsometer software library. We assumed the ITO could be treated as a single bulk layer with a surface roughness layer modeled as an EMA mixture with vacuum, as suggested by Lohner et al \cite{lohner}. 

After reduction, we assumed the film could be approximated by an upper layer with optical constants changed by reduction on top of a lower layer that was unchanged. We modeled the lower layer by allowing the original ITO thickness to vary while its optical constants remained the same. We replaced the surface roughness layer with an additional bulk layer of ITO which we allowed to vary in thickness and optical parameters to model the upper layer of reduced nanoparticles. In total, we used eleven parameters to fit the reduced ITO: underlying layer thickness, nanoparticle layer thickness, the strengths of the two Lorentzian oscillators $f^1$ and $f^2$, the broadening of the two Lorentzian oscillators $\Gamma^1$ and $\Gamma^2$, the resonant frequencies of the two Lorentzian oscillators $E^1_0$ and $E^2_0$, the Drude oscillator resonant frequency $E_p$, the Drude oscillator broadness $E_\Gamma$, and the high-frequency dielectric constant $\epsilon_\infty$.

\begin{table}[ht!]
\caption{Optical Layer Thicknesses}
Thicknesses resulting from fitting the Drude-Lorentz model for the reduced ITO bilayer.
\begin{center}
    \begin{tabular}{|c|c|c|}
        \hline
         Sample ID & \thead{Bulk \\ Thickness [nm]} & \thead{Nanoparticle \\ Thickness [nm]} \\
         \hline
         Non-reduced & 136 - 138 & - \\
         60 s (a) & 121 & 48 \\
         60 s (b) & 120 & 50 \\
         90 s (a) & 120 & 59 \\
         90 s (b) & 121 & 69 \\
         120 s (a) & 112 & 71 \\
         120 s (b) & 120 & 62 \\
         150 s (a) & 112 & 73 \\
         150 s (b) & 119 & 77 \\
         180 s (a) & 113 & 79 \\
         180 s (b) & 117 & 91 \\
         \hline
    \end{tabular}
\end{center}
\label{tab:thickness}
\end{table}

\begin{table}[ht!]
\caption{Optical Constants}
Optical constants resulting from fitting the Drude-Lorentz model for the reduced ITO bilayer.
\begin{center}
    \begin{tabular}{|c|c|c|c|c|c|c|c|c|c|}
        \hline
         \thead{Sample ID} & $E^1_0$ [eV] & $f^1$ & $\Gamma^1$ [eV] & $E^2_0$ [eV] & $f^2$ & $\Gamma^2$ [eV] & $E_p$ [eV] & $E_\Gamma$ [eV] & $\epsilon_\infty$ \\
         \hline
         Non-reduced & 3.75 & 0.16 & 0.62 & 4.30 & 0.44 & 0.24 & 1.10 & 0.23 & 3.50 \\
         60 s (a) & 3.28 & 0.14 & 0.93 & 4.30 & 0.35 & 0.55 & 0.66 & 0.44 & 1.44 \\
         60 s (b) & 3.24 & 0.27 & 1.36 & 4.22 & 0.22 & 0.32 & 0.84 & 0.43 & 1.47 \\
         90 s (a) & 3.36 & 0.26 & 1.22 & 4.29 & 0.25 & 0.54 & 0.67 & 0.45 & 1.42 \\
         90 s (b) & 3.14 & 0.30 & 1.42 & 4.19 & 0.18 & 0.43 & 0.74 & 0.44 & 1.53 \\
         120 s (a) & 3.45 & 0.49 & 1.13 & 4.31 & 0.15 & 0.58 & 0.60 & 0.28 & 1.57 \\
         120 s (b) & 3.40 & 0.40 & 1.34 & 4.37 & 0.27 & 0.67 & 0.68 & 0.43 & 1.40 \\
         150 s (a) & 3.36 & 0.47 & 1.14 & 4.23 & 0.17 & 0.73 & 0.68 & 0.32 & 1.63 \\
         150 s (b) & 3.21 & 0.36 & 1.41 & 4.25 & 0.21 & 0.63 & 0.70 & 0.40 & 1.44 \\
         180 s (a) & 3.22 & 0.44 & 1.19 & 4.19 & 0.25 & 0.93 & 0.63 & 0.31 & 1.58 \\
         180 s (b) & 2.98 & 0.42 & 1.37 & 4.18 & 0.30 & 0.96 & 0.69 & 0.36 & 1.47 \\
         \hline
    \end{tabular}
\end{center}
\label{tab:optical}
\end{table}
\clearpage

\bibliographystyle{unsrt}
\bibliography{ito-bib}

\begin{thebibliography}{10}

\bibitem{allman_near-infrared_2015}
M.~S. Allman, V.~B. Verma, M.~Stevens, T.~Gerrits, R.~D. Horansky, A.~E. Lita,
  F.~Marsili, A.~Beyer, M.~D. Shaw, D.~Kumor, R.~Mirin, and S.~W. Nam.
\newblock A near-infrared 64-pixel superconducting nanowire single photon
  detector array with integrated multiplexed readout.
\newblock {\em Appl. Phys. Lett.}, 106(19):192601, May 2015.
\newblock Publisher: American Institute of Physics.

\bibitem{holzgrafe_cavity_2020}
Jeffrey Holzgrafe, Neil Sinclair, Neil Sinclair, Di~Zhu, Di~Zhu, Amirhassan
  Shams-Ansari, Marco Colangelo, Yaowen Hu, Yaowen Hu, Mian Zhang, Mian Zhang,
  Karl~K. Berggren, and Marko Lončar.
\newblock Cavity electro-optics in thin-film lithium niobate for efficient
  microwave-to-optical transduction.
\newblock {\em Optica, OPTICA}, 7(12):1714--1720, December 2020.
\newblock Publisher: Optica Publishing Group.

\bibitem{rueda_efficient_2016}
Alfredo Rueda, Florian Sedlmeir, Michele~C. Collodo, Ulrich Vogl, Birgit
  Stiller, Gerhard Schunk, Dmitry~V. Strekalov, Christoph Marquardt,
  Johannes~M. Fink, Oskar Painter, Gerd Leuchs, and Harald G.~L. Schwefel.
\newblock Efficient microwave to optical photon conversion: an electro-optical
  realization.
\newblock {\em Optica, OPTICA}, 3(6):597--604, June 2016.
\newblock Publisher: Optica Publishing Group.

\bibitem{budoyo_effects_2016}
R.~P. Budoyo, J.~B. Hertzberg, C.~J. Ballard, K.~D. Voigt, Z.~Kim, J.~R.
  Anderson, C.~J. Lobb, and F.~C. Wellstood.
\newblock Effects of nonequilibrium quasiparticles in a thin-film
  superconducting microwave resonator under optical illumination.
\newblock {\em Phys. Rev. B}, 93(2):024514, January 2016.
\newblock Publisher: American Physical Society.

\bibitem{barends_minimizing_2011}
R.~Barends, J.~Wenner, M.~Lenander, Y.~Chen, R.~C. Bialczak, J.~Kelly,
  E.~Lucero, P.~O’Malley, M.~Mariantoni, D.~Sank, H.~Wang, T.~C. White,
  Y.~Yin, J.~Zhao, A.~N. Cleland, John~M. Martinis, and J.~J.~A. Baselmans.
\newblock Minimizing quasiparticle generation from stray infrared light in
  superconducting quantum circuits.
\newblock {\em Appl. Phys. Lett.}, 99(11):113507, September 2011.
\newblock Publisher: American Institute of Physics.

\bibitem{mirhosseini_superconducting_2020}
Mohammad Mirhosseini, Alp Sipahigil, Mahmoud Kalaee, and Oskar Painter.
\newblock Superconducting qubit to optical photon transduction.
\newblock {\em Nature}, 588(7839):599--603, December 2020.
\newblock Number: 7839 Publisher: Nature Publishing Group.

\bibitem{bousquet_etude_1957}
P~Bousquet.
\newblock Etude theorique des proprietes optiques des couches minces
  transparentes.
\newblock {\em Annales de physique}, 13:5--15, 1957.

\bibitem{leksina_optical_nodate}
E~Leksina, G~P Motulevich, and A~A Shubin.
\newblock The optical properties of niobium.
\newblock {\em Soviet Physics Jetp}, 29:8, July 1968.

\bibitem{bernardo_progress_2021}
Gabriel Bernardo, Tânia Lopes, David~G. Lidzey, and Adélio Mendes.
\newblock Progress in {Upscaling} {Organic} {Photovoltaic} {Devices}.
\newblock {\em Advanced Energy Materials}, 11(23):2100342, 2021.
\newblock \_eprint:
  https://onlinelibrary.wiley.com/doi/pdf/10.1002/aenm.202100342.

\bibitem{betz_thin_2006}
U.~Betz, M.~Kharrazi~Olsson, J.~Marthy, M.~F. Escolá, and F.~Atamny.
\newblock Thin films engineering of indium tin oxide: {Large} area flat panel
  displays application.
\newblock {\em Surface and Coatings Technology}, 200(20):5751--5759, May 2006.

\bibitem{aydin_indium_2017}
Elif~Burcu Aydın and Mustafa~Kemal Sezgintürk.
\newblock Indium tin oxide ({ITO}): {A} promising material in biosensing
  technology.
\newblock {\em TrAC Trends in Analytical Chemistry}, 97:309--315, December
  2017.

\bibitem{edwards_basic_2004}
P.~P. Edwards, A.~Porch, M.~O. Jones, D.~V. Morgan, and R.~M. Perks.
\newblock Basic materials physics of transparent conducting oxides.
\newblock {\em Dalton Trans.}, (19):2995, 2004.

\bibitem{ohyama_weak_1985}
Tyuzi Ohyama, Minoru Okamoto, and Eizo Otsuka.
\newblock Weak {Localization} and {Correlation} {Effects} in
  {Indium}-{Tin}-{Oxide} {Films}. {II}. {Two}-to-{Three} {Dimensional}
  {Transition} and {Competition} between {Localization} and
  {Superconductivity}.
\newblock {\em J. Phys. Soc. Jpn.}, 54(3):1041--1053, March 1985.
\newblock Publisher: The Physical Society of Japan.

\bibitem{mori_superconductivity_1993}
Natsuki Mori.
\newblock Superconductivity in transparent {Sn}‐doped {In2O3} films.
\newblock {\em Journal of Applied Physics}, 73(3):1327--1338, February 1993.
\newblock Publisher: American Institute of Physics.

\bibitem{chiu_four-probe_2009}
Shao-Pin Chiu, Hui-Fang Chung, Yong-Han Lin, Ji-Jung Kai, Fu-Rong Chen, and
  Juhn-Jong Lin.
\newblock Four-probe electrical-transport measurements on single indium tin
  oxide nanowires between 1.5 and 300 {K}.
\newblock {\em Nanotechnology}, 20(10):105203, March 2009.

\bibitem{aliev_reversible_2012}
Ali~E. Aliev, Ka~Xiong, Kyeongjae Cho, and M.~B. Salamon.
\newblock Reversible superconductivity in electrochromic indium-tin oxide
  films.
\newblock {\em Appl. Phys. Lett.}, 101(25):252603, December 2012.
\newblock Publisher: American Institute of Physics.

\bibitem{bressers_electrochromic_1998}
Peter M. M.~C. Bressers and Eric~A. Meulenkamp.
\newblock The {Electrochromic} {Behavior} of {Indium} {Tin} {Oxide} in
  {Propylene} {Carbonate} {Solutions}.
\newblock {\em J. Electrochem. Soc.}, 145(7):2225--2231, July 1998.

\bibitem{liu_important_2015}
Liang Liu, Shai Yellinek, Ido Valdinger, Ariela Donval, and Daniel Mandler.
\newblock Important {Implications} of the {Electrochemical} {Reduction} of
  {ITO}.
\newblock {\em Electrochimica Acta}, 176:1374--1381, September 2015.

\bibitem{batson_reduced_2022}
Emma Batson.
\newblock Reduced {Indium} {Tin} {Oxide} as a {Transparent} {Superconductor}.
\newblock Master's thesis, Massachusetts Institute of Technology, May 2022.

\bibitem{noauthor_indium_nodate}
Indium tin oxide coated glass slide, square surface resistivity 8-12ohm/sq
  50926-11-9.

\bibitem{baum_determination_2013}
M.~Baum, I.~Alexeev, M.~Latzel, S.~H. Christiansen, and M.~Schmidt.
\newblock Determination of the effective refractive index of nanoparticulate
  {ITO} layers.
\newblock {\em Opt. Express, OE}, 21(19):22754--22761, September 2013.
\newblock Publisher: Optica Publishing Group.

\bibitem{koonce_superconducting_1967}
C.~S. Koonce, Marvin~L. Cohen, J.~F. Schooley, W.~R. Hosler, and E.~R.
  Pfeiffer.
\newblock Superconducting {Transition} {Temperatures} of {Semiconducting}
  {SrTiO3}.
\newblock {\em Phys. Rev.}, 163(2):380--390, November 1967.
\newblock Publisher: American Physical Society.

\bibitem{ueno_discovery_2011}
K.~Ueno, S.~Nakamura, H.~Shimotani, H.~T. Yuan, N.~Kimura, T.~Nojima, H.~Aoki,
  Y.~Iwasa, and M.~Kawasaki.
\newblock Discovery of superconductivity in {KTaO3} by electrostatic carrier
  doping.
\newblock {\em Nature Nanotech}, 6(7), July 2011.

\bibitem{das_superconducting_2015}
Tanmoy Das and Kapildeb Dolui.
\newblock Superconducting dome in {MoS} 2 and {TiSe} 2 generated by
  quasiparticle-phonon coupling.
\newblock {\em Phys. Rev. B}, 91(9):094510, March 2015.

\bibitem{yu_superconducting_2020}
Xiang-Long Yu and Jiansheng Wu.
\newblock Superconducting dome driven by intervalley phonon scattering in
  monolayer {MoS} $_{\textrm{2}}$.
\newblock {\em New J. Phys.}, 22(1):013015, January 2020.

\bibitem{pearl_vortex_nodate}
Judea Pearl.
\newblock {\em Vortex {Theory} of {Superconductive} {Memories}}.
\newblock Ph.{D}., Polytechnic University, United States -- New York.
\newblock ISBN: 9798658009575.

\bibitem{kowal_scale_2008}
D.~Kowal and Z.~Ovadyahu.
\newblock Scale dependent superconductor–insulator transition.
\newblock {\em Physica C: Superconductivity}, 468(4):322--325, February 2008.

\bibitem{matveeva_electrochemistry_2005}
Eugenia Matveeva.
\newblock Electrochemistry of the {Indium}-{Tin} {Oxide} {Electrode} in 1 {M}
  {NaOH} {Electrolyte}.
\newblock {\em J. Electrochem. Soc.}, 152(9):H138, 2005.

\bibitem{spada_role_2013}
E.~R. Spada, F.~R. de~Paula, C.~C. Plá~Cid, G.~Candiotto, R.~M. Faria, and
  M.~L. Sartorelli.
\newblock Role of acidic and basic electrolytes on the structure and morphology
  of cathodically reduced indium tin oxide ({ITO}) substrates.
\newblock {\em Electrochimica Acta}, 108:520--524, October 2013.

\bibitem{bouden_multifunctional_2016}
Sarra Bouden, Antoine Dahi, Fanny Hauquier, Hyacinthe Randriamahazaka, and
  Jalal Ghilane.
\newblock Multifunctional {Indium} {Tin} {Oxide} {Electrode} {Generated} by
  {Unusual} {Surface} {Modification}.
\newblock {\em Sci Rep}, 6(1), November 2016.

\bibitem{zhang_metal_2020}
Chao Zhang, Guifang Liu, Xin Geng, Kaidi Wu, and Marc Debliquy.
\newblock Metal oxide semiconductors with highly concentrated oxygen vacancies
  for gas sensing materials: {A} review.
\newblock {\em Sensors and Actuators A: Physical}, 309:112026, July 2020.

\bibitem{kanehara_indium_2009}
Masayuki Kanehara, Hayato Koike, Taizo Yoshinaga, and Toshiharu Teranishi.
\newblock Indium {Tin} {Oxide} {Nanoparticles} with {Compositionally} {Tunable}
  {Surface} {Plasmon} {Resonance} {Frequencies} in the {Near}-{IR} {Region}.
\newblock {\em J. Am. Chem. Soc.}, 131(49):17736--17737, December 2009.
\newblock Publisher: American Chemical Society.

\bibitem{ohsawa_origin_2020}
Takeo Ohsawa, Naoomi Yamada, Akichika Kumatani, Yoshitaka Takagi, Tohru Suzuki,
  Ryota Shimizu, Susumu Shiraki, Tsutomu Nojima, and Taro Hitosugi.
\newblock Origin of {Optical} {Transparency} in a {Transparent}
  {Superconductor} {LiTi2O4}.
\newblock {\em ACS Applied Electronic Materials}, 2(2):517--522, February 2020.
\newblock Publisher: American Chemical Society.

\bibitem{lohner}
Tivadar Lohner, K.~Jagadeesh Kumar, Péter Petrik, Aryasomayajula Subrahmanyam,
  and István Bársony.
\newblock Optical analysis of room temperature magnetron sputtered ito films by
  reflectometry and spectroscopic ellipsometry.
\newblock {\em Journal of Materials Research}, 29:1528--1536, 2014.

\end{thebibliography}

\end{document}